\documentclass[conference]{IEEEtran}
\IEEEoverridecommandlockouts
\usepackage{cite}
\usepackage{amsmath,amssymb,amsfonts}
\usepackage{algorithmic}
\usepackage{graphicx}
\usepackage{stfloats}
\usepackage{textcomp}
\usepackage[bookmarks=false]{hyperref}
\hypersetup{
    colorlinks = true,
    citecolor  = blue,
    linkcolor  = blue,
    urlcolor   = blue,
}
\usepackage{colortbl}
\usepackage[table]{xcolor}
\usepackage{multirow}
\usepackage{booktabs}
\usepackage{cleveref}
\usepackage{subcaption}
\def\equationautorefname~#1\null{Eq.~(#1)\null}
\def\figureautorefname~#1\null{Fig.~#1\null}
\def\tableautorefname~#1\null{Tab.~#1\null}
\def\definitionautorefname~#1\null{Def.~#1\null}
\def\sectionautorefname~#1\null{Sect.~#1\null}
\def\subsectionautorefname~#1\null{Sect.~#1\null}
\def\subsubsectionautorefname~#1\null{Sect.~#1\null}

\def\BibTeX{{\rm B\kern-.05em{\sc i\kern-.025em b}\kern-.08em
    T\kern-.1667em\lower.7ex\hbox{E}\kern-.125emX}}

\title{\raisebox{-4pt}{\includegraphics[height=28pt]{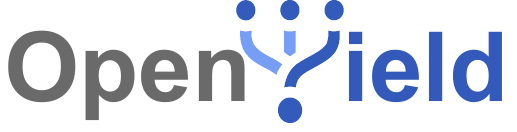}}: An {Open}-Source SRAM {Yield} Analysis and Optimization Benchmark Suite
\vspace{-10pt}
} 

\author{
\IEEEauthorblockN{Shan Shen$^{1,\#}$, Xingyang Li$^{2,\#}$, Zhuohua Liu$^{2,\#}$, Junhao Ma$^{1}$, Yikai Wang$^{1}$, Yiheng Wu$^{1}$,  \\ Yuquan Sun$^{2}$ and Wei W. Xing$^{3,*}$}

\IEEEauthorblockA{$^{1}$\textit{Nanjing University of Science and Technology, Nanjing, 210094, China} \\
$^{2}$\textit{Beihang University, Beijing, 100191, China} \\
$^{3}$\textit{University of Sheffield, Sheffield, S102TN, United Kingdom}}

\IEEEauthorblockA{Email: \{shanshen, marvin660, wangyikai, wuyiheng\}@njust.edu.cn, \\ 
\{lixingyang, zhuohualiu, sunyq\}@buaa.edu.cn,  w.xing@sheffield.ac.uk}

\thanks{
This work is supported by the National Natural Science Foundation of China (NSFC) under grant No. 62204141 and the Fundamental Research Funds for the Central Universities under grant No. 30925010605. 

$^\#$ Equal contribution. $^*$ Corresponding author. }
\vspace{-20pt}
}

\begin{document}

\maketitle

\begin{abstract}

Static Random-Access Memory (SRAM) yield analysis is essential for semiconductor innovation, yet research progress faces a critical challenge: the large gap between simplified academic models and the complexities observed in practice. The lack of open, higher-fidelity benchmarks has hindered reproducibility and transferability, as promising academic techniques often fail to carry over to more realistic settings. We present \textit{OpenYield}, an open-source ecosystem that aims to narrow this gap through three contributions: (i) An SRAM circuit generator that explicitly incorporates second-order effects (interconnect/line parasitics, inter-cell leakage coupling, and peripheral-circuit variations) that are commonly omitted in academic studies. (ii) A standardized evaluation platform with a simple interface and baseline yield-analysis implementations to enable fair comparisons and reproducible research on these higher-fidelity circuits. (iii) An optimization platform for transistor-level sizing under these models, supporting reproducible studies of robustness/efficiency trade-offs.
OpenYield aims to foster more reproducible and transferable progress in SRAM-yield research. The framework is publicly available at \href{https://github.com/ShenShan123/OpenYield}{OpenYield:URL}.

\end{abstract}

\section{Introduction}
\label{sec:introduction}

Static Random-Access Memory (SRAM) has emerged as the dominant on-chip memory technology in modern integrated circuits, consuming up to 70\% of die area in advanced AI accelerators and 60\% in typical mobile System-on-Chips (SoCs) \cite{chandrakasan2019sram}. As semiconductor manufacturing pushes into sub-5nm nodes, the economic stakes of SRAM yield have become paramount—a mere 1\% yield loss in high-volume production translates to millions in revenue impact, making accurate yield prediction not just a technical challenge but a critical business imperative for the \$580 billion semiconductor industry \cite{agarwal2008statistical}.

The computational challenge of SRAM yield analysis is daunting. Modern memory arrays require failure rates below $10^{-9}$ to be viable, yet directly observing such rare events would require over $10^{10}$ Monte Carlo simulations—computationally infeasible even with today's massive computing resources \cite{kanj2006mixture}. This fundamental limitation has spurred decades of research into more efficient techniques, from variance reduction methods like importance sampling \cite{dolecek2008breaking} and statistical blockade \cite{singhee2008practical}, to modern machine learning approaches \cite{bouhlila2024machine, yoon2022machine} and surrogate modeling techniques \cite{wang2017efficient, yao2014efficient}. Each promises to dramatically reduce simulation costs while maintaining accuracy.


Yet despite this rich algorithmic innovation, a troubling pattern has emerged: techniques that show exceptional promise in academic publications often fail to transfer to production designs. Our comprehensive review of 45 recent SRAM yield analysis papers reveals the root cause—over 85\% rely on dramatically oversimplified models that omit the very physical effects in real circuits. This systematic disconnect between academic models and production reality has created what we term a ``\textit{reproducibility crisis}'' in SRAM yield analysis.

The gap between academic models and industrial reality is qualitative, not merely quantitative. Real SRAM arrays exhibit complex phenomena that fundamentally alter behavior: parasitic resistances and capacitances slow bitline discharge by 4-6$\times$ \cite{shen2024deep}; leakage from hundreds of unselected cells degrades read margins by 12\% \cite{shen2024ultra8t}; sense amplifier offset variations of 15-30mV shift failure boundaries unpredictably \cite{abu2011characterization, shen2022single}; and position-dependent IR drops create systematic variations across arrays \cite{shen2024sram}. When combined, these effects create failure modes that simply cannot exist in simplified models.
This modeling gap has severe practical consequences. The semiconductor industry spends \$2-3 billion annually on design respins, with yield issues driving 30\% of these costs. More critically, the disconnect undermines research itself: algorithms showing similar performance on simplified benchmarks may differ by orders of magnitude on real circuits; optimization techniques that appear optimal in studies fail catastrophically in practice; and without standardized, higher-fidelity benchmarks, fair comparison of approaches becomes impossible.

To address this systemic challenge, we present \textbf{OpenYield}: an open-source ecosystem for SRAM yield analysis and optimization that aims to narrow the gap between research circuits and real SRAM modules in the on-chip systems. Our goal is to make higher-fidelity, open benchmarks more accessible, enabling the academic community to develop and validate techniques on circuits that incorporate key second-order effects observed in practice.
OpenYield makes three core novelties:

\begin{itemize}
    \item \textbf{A Python-based, scalable SRAM generator} that configures any-sized array and explicitly models key wire parasitics, inter-cell leakage, and peripheral circuit variations—effects often omitted in academic studies. Validation shows models without parasitics underestimate read delay and power by {3.7$\times$ and 2.5$\times$}; models without peripherals underrate by {2.3$\times$ and 1.28$\times$}.
    \item \textbf{A Standardized Yield Analysis Platform} with reference implementations of baseline algorithms (Monte Carlo and importance sampling variants) operating on circuits that include these effects, enabling reproducible benchmarking across different SRAM configurations.
    \item \textbf{A Standardized SRAM Optimization Platform} with five state-of-the-art optimization algorithms for transistor sizing under these models, enabling fair comparison and demonstrating up to 65\% SNM improvements and 15\% area reduction over baseline designs.
\end{itemize}


OpenYield aims to transform SRAM yield analysis/optimization research by providing higher-fidelity test circuits and baseline methods in an open-source platform. Like ImageNet  \cite{Deng2009imagenet} for computer vision or SPEC for computer architecture \cite{Henning2006spec}, it establishes standardized benchmarks that enable reproducible research and new technical directions. The complete framework, with documentation and examples, will be publicly available to maximize adoption.

\section{Background}
\label{sec:background}

\subsection{Physical Origins of the SRAM Yield Challenge}
\label{ssec:physical_origins}

Process variations in semiconductor manufacturing create the fundamental yield challenge in SRAM design. As transistors shrink to atomic scales, random dopant fluctuation becomes inevitable—a 20nm transistor contains only ~100 dopant atoms in its channel, making each atom's position statistically significant \cite{asenov2003random}. These atomic-level variations manifest as threshold voltage ($V_{TH}$) mismatch between transistors, with standard deviations following Pelgrom's law: $\sigma_{V_{TH}} = A_{vt}/\sqrt{WL}$, where $A_{vt}$ ranges from 2-4 mV·µm in modern processes \cite{pelgrom1989matching}.

SRAM cells amplify these variations through their ratioed design. Read stability requires the pull-down transistor (PD) to overpower the access (PG) transistor, preventing the stored node from flipping. Write ability demands the opposite—the PG transistor must overpower the pull-up (PU). These conflicting requirements create narrow design margins that variations easily violate. A 30mV $V_{TH}$ shift well within 3$\sigma$ can push a marginal cell into failure.
The statistics compound mercilessly. A 1Mb array contains over 6 million transistors. Even with 6$\sigma$ process control, several thousand transistors will exhibit extreme variations. Since a single failing cell renders the entire array unusable, achieving acceptable yield requires individual cell failure rates below $10^{-7}$, deep in the statistical tail, where rare events dominate.

\subsection{Formulation of Yield Analysis and SRAM Optimization}
\label{ssec:yield_formulation}

Yield analysis quantifies the probability that a manufactured array functions correctly despite process variations. For an array with $N$ identical and independent cells, where each cell has a failure probability of $P_{f}$, the overall yield $Y$ can be expressed as $Y=(1 - P_{f})^N$. For small failure probabilities, this can be approximated as $Y\approx \exp(-N \cdot P_{f})$. This simple relationship shows how even a small per-cell failure probability can dramatically impact the yield of large arrays.

This exponential relationship drives the stringent requirements. Achieving 90\% yield for a 1Mb array demands $P_{f} < 1.05 \times 10^{-7}$.
Monte Carlo simulation provides the reference method for estimating $P_{f}$. Generate $M$ samples from the process distribution, simulate each, calculate failures as $P_{f} = N_{f} / M$. For $P_{f} = 10^{-7}$, this demands $M \approx 10^9$ simulations.

SRAM optimization navigates competing objectives under yield constraints, formulated as 
\begin{equation}
\begin{split}
& \mathop{\arg\min}\limits_{\mathbf{x}} \, \left [f_1(\mathbf{x}), f_2(\mathbf{x}), \ldots, f_m(\mathbf{x})\right ], \\
& \text{s.t.}~T \leq t_{spec}, 
\end{split}
\end{equation}
where $\mathbf{x}$ represents transistors' widths (\texttt{W}) and lengths (\texttt{L}) in a SRAM cell, $f_j(\mathbf{x})$ are objectives (e.g., power, area, noise margin), and $T$ can be yield-related timing constraints. Critically, these relationships change dramatically based on which physical effects the model includes.

\subsection{The Modeling Gap: From Simplification to Crisis}
\label{ssec:modeling_gap}

Academic SRAM research suffers from systematic oversimplification. 
Real arrays exhibit failure modes that simplified models cannot capture:
\begin{itemize}
\item \textbf{Distributed parasitics} create position-dependent behavior. A 256-row bitline presents $>200$fF capacitance and $>1k\Omega$ resistance, not as lumped elements but distributed along the wire. Cells at the array's far end see 4-6$\times$ slower discharge rates \cite{shen2024deep}. Simple RC models miss the incomplete swing and non-monotonic delay that distributed effects create.

\item \textbf{Inter-cell coupling} introduces data-pattern dependencies. With 255 unselected cells per bitline, cumulative leakage reaches microamps—comparable to the selected cell's read current at low voltages \cite{shen2024ultra8t}. The resulting margin degradation depends on stored data patterns, creating failures invisible to single-cell analysis.

\item \textbf{Peripheral circuits} often dominate yield loss. Sense amplifier (SA) offset voltage ($\sigma_{V_{\text{OS}}} \approx$ 20mV) directly subtracts from read margin \cite{abu2011characterization, shen2022single}. Wordline driver skew creates systematic timing variations across columns. These effects require modeling the complete SRAM array, not just the core cell.

\item \textbf{Other second-order effects} include: (i) layout-aware spatial/systematic variation (pattern density, well proximity, strain/stress, orientation); (ii) supply-network effects (static IR drop and dynamic supply bounce on VDD/VSS); and (iii) temperature and aging (self-heating/thermal gradients, BTI/HCI-induced shifts).
\end{itemize}

We prioritize the first three effects because, in mainstream 2D CMOS SRAMs and the low-voltage operating points we study, these are the dominant contributors to read/write margin loss and sense errors. Other effects remain important and are compatible with our framework, and we leave the integration of them to future work.

\subsection{SOTA SRAM Tools and Their Limitations}
\label{ssec:existing_tools}
Current SRAM tools serve specific niches but fail to address yield analysis needs comprehensively.
OpenRAM \cite{guthaus2016openram} compiles complete SRAMs from specification to GDS with diverse configurations and manufacturable layouts, but lacks variation modeling and yield analysis infrastructure.
SRAM22 \cite{SRAM22} provides automated embedded SRAM generation with compact layouts, but similarly offers no statistical analysis or yield optimization features.
AnalogGym \cite{li2024analoggym} benchmarks general analog synthesis algorithms but cannot capture SRAM-specific phenomena like massive parallelism, statistical extremes, and critical layout dependencies.

Commercial tools accurately model these effects but remain inaccessible due to proprietary licenses, black-box implementations, and high costs that exclude academic research.
OpenYield addresses this gap by providing open-source models that incorporate many second-order effects beyond common academic simplifications, enabling reproducible research and more informative comparisons.

\begin{figure}[!t]
\vspace{-4pt}
\setlength{\abovecaptionskip}{0pt}
    \centering
    \includegraphics[width=0.95\linewidth]{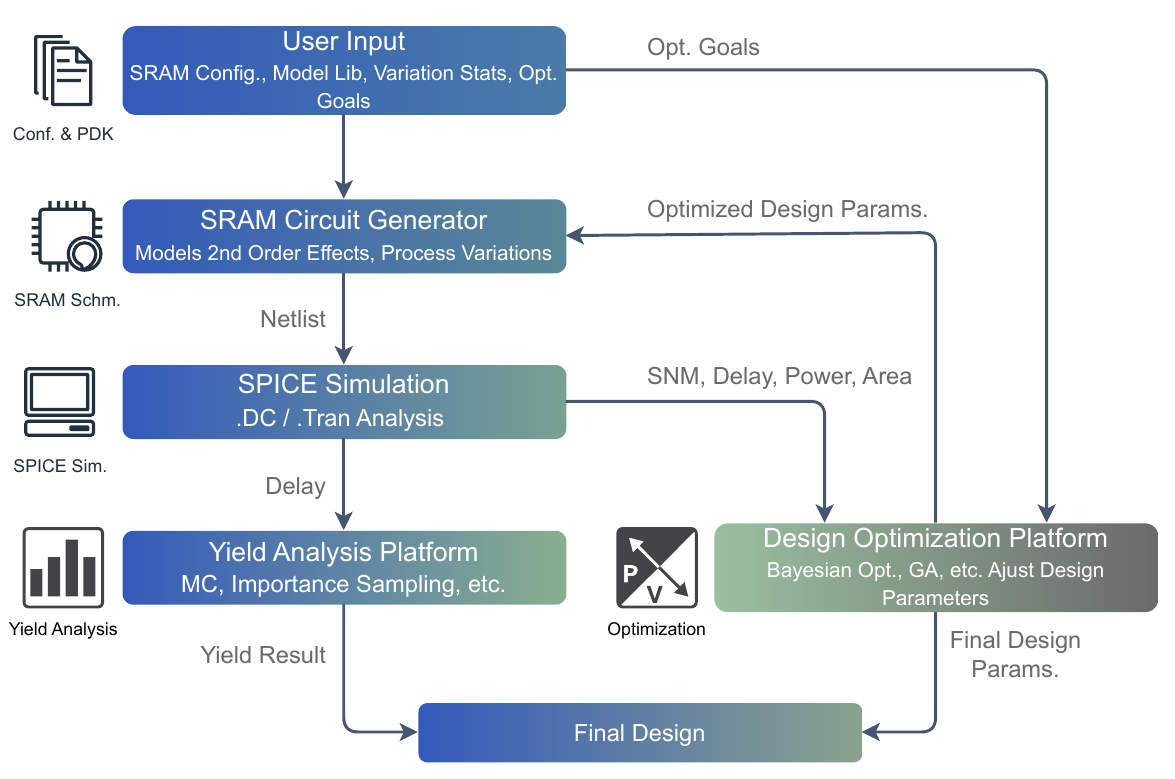}
    \caption{Overview of OpenYield's ecosystem showing the circuit generator, baseline yield analysis algorithms, and transistor-sizing optimization framework.}
    \label{fig:openyield_architecture}
    \vspace{-4pt}
\end{figure}

\section{\raisebox{-2pt}{\includegraphics[height=11.9pt]{illu/openyield_all-logo-hori.drawio.pdf}}: Design and Implementation}
\label{sec:methodology}

The OpenYield ecosystem follows a modular architecture depicted in \autoref{fig:openyield_architecture}. The core \textit{SRAM Circuit Generator} produces PySpice-based netlists \cite{PySpice} that incorporate user configurations and relevant physical effects. These netlists are analyzed by an engine using open-source SPICE tools like Xyce \cite{keiter2013xyce}.

Two principal blocks leverage the simulation insights: \textit{Yield Analysis Algorithms} (\autoref{ssec:baseline_algorithms}) implement statistical techniques for yield estimation and failure probability assessment, while the \textit{Design Optimization Framework} (\autoref{ssec:optimization}) automatically adjusts design parameters (e.g., bitcell transistor sizes) to enhance SRAM robustness and performance.

Users interact with OpenYield by providing comprehensive inputs including SRAM configurations (arbitrary array dimensions, memory cell types from 6T standard with 8T/10T provisions), technology characteristics (currently FreePDK45 \cite{FreePDK45} support), design parameters, process variation statistics (local intra-die random variations applied independently to each transistor using Gaussian distributions and global process corners), and optimization goals for Monte Carlo simulations. The platform delivers multiple outputs: SPICE netlists (\texttt{.sp} format) ensuring broad simulator compatibility, detailed yield reports, failure probability estimates, characterized performance metrics, and optimized design parameters. Monte Carlo parameterization can be embedded directly in model cards or exported as data tables for importance sampling-based yield analysis algorithms. The platform follows open-source principles, encouraging community contributions and extensions.

\subsection{Hierarchical SRAM Generator}
\label{ssec:generator}

The cornerstone of OpenYield is its novel SRAM circuit generator, engineered to produce unified benchmark circuits that transcend common academic simplifications and approach SRAM complexities observed in practice.

\subsubsection{Circuit Architecture}

\begin{figure}[!t] 
\centering
\begin{subfigure}[b]{0.49\linewidth} 
\includegraphics[width=\textwidth]{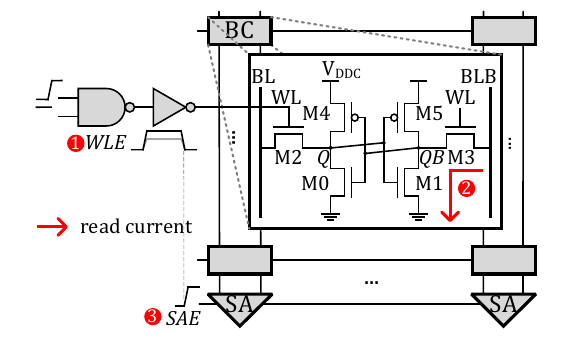} 
\caption{Read `1': \texttt{WL} high, \texttt{Q} high, \texttt{QB} low, \texttt{BLB} discharges.}
\label{fig:sram_read_op}
\end{subfigure}
\hfill 
\begin{subfigure}[b]{0.49\linewidth}
\includegraphics[width=\textwidth]{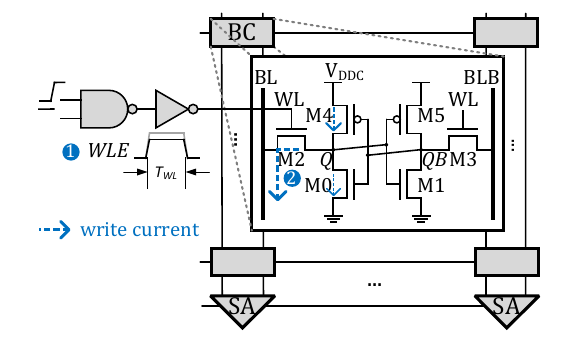} 
\caption{Write '0' to '1': \texttt{WL} high, \texttt{BL} low, \texttt{BLB} high.}
\label{fig:sram_write_op}
\end{subfigure}
\caption{Schematics of a 6T SRAM cell under read and write operations.}
\label{fig:sram_ops} 
\vspace{-4pt}
\end{figure}

As illustrated by \autoref{fig:sram_ops}, the 6T SRAM cell comprises cross-coupled inverters (M0-M1, M4-M5) for storage and PG (M2-M3) for read/write control. Process variations, particularly random dopant fluctuation, create threshold voltage mismatches that drive yield loss through specific failure mechanisms:

\begin{itemize}
\item \textbf{Read Upsets} occur when bitline \texttt{BLB} discharge through M3-M1 causes node $\text{V}(\texttt{QB})>\text{V}_\text{TRIP}$ of M4-M0 inverter, destructively flipping the cell state. Read Static Noise Margin (RSNM) quantifies the cell's immunity to this—insufficient RSNM (typically $<$50mV) directly causes the failure.

\item \textbf{Write failures} happen when M2 cannot pull node \texttt{Q} below the M5-M1 trip point against M4's pull-up action. Write Static Noise Margin (WSNM) measures the cell's ability to flip during write operations—inadequate WSNM ($<$50mV) prevents successful data writing.

\item \textbf{Hold failures} occur when the cell loses stored data due to leakage currents or \texttt{VDD} droops in standby mode, when both bitlines float and the wordline is inactive. Hold Static Noise Margin (HSNM) characterizes data retention stability—insufficient HSNM ($<$100mV) leads to spontaneous data corruption during idle periods.

\item \textbf{Access time failures} result when the bitline differential cannot overcome the $V_{\text{OS}}$ within the allocated timing window. Both read delay ($T_{\text{READ}}$, including time to develop sufficient bitline differential) and write delay ($T_{\text{WRITE}}$, including time to flip the cell state) must meet timing constraints—violations cause functional failures regardless of noise margins.
\end{itemize}

All these failure mechanisms can be simulated and analyzed through OpenYield. The generator constructs complete array hierarchies including: core cells in configurable row/column arrangements, wordline drivers sized for distributed loads, column multiplexers with shared sense amplifiers, precharge/write driver circuits, sense amplifiers with statistical offset distributions, and configurable control logic. This comprehensive approach distinguishes OpenYield from existing academic generators that produce idealized circuits.

\subsubsection{Second-Order Physical Effects Modeling}

OpenYield integrates critical physical effects that profoundly influence yield in advanced technologies, as listed below.
\begin{itemize}
\item \textbf{Parasitic Network Generation:} Distributed $\pi$-shaped RC segments model vital conductive paths (bitlines, wordlines) rather than oversimplified lumped capacitances. Users can specify resistance/capacitance values and segment numbers for each subcircuit, with defaults based on typical SRAM layout topologies.

\item \textbf{Inter-Cell Leakage Coupling:} Models various leakage mechanisms (subthreshold, gate leakage) to simulate neighboring cell and peripherals impacts on actively accessed cells—particularly significant under low \texttt{VDD} conditions and deeper SRAM designs.

\item \textbf{Peripheral Circuit Variation:} Instantiates statistical models for decoders (capturing stage-dependent delay on the critical access path and output slew that feeds into the wordline drivers), wordline drivers (with delay variations and skews, and sensitivity to decoder output slew), sense amplifiers (with systematic/random offset voltage distributions), and precharge/write drivers (with voltage level and timing variations). These peripheral variations are modeled conjunctively with core cell variations for holistic yield assessment.
\end{itemize}

Other second-order effects discussed in \autoref{ssec:modeling_gap}—including power-grid IR drop, layout-aware spatial/systematic variation driven by device geometry, long-term aging (BTI/HCI), and thermal gradients—are not modeled by default. The framework provides hooks to incorporate them: IR drop can be incorporated by instantiating additional RC elements on the supply rails to model impedance; layout-aware variation can be parameterized via physical design attributes (e.g., diffusion areas/perimeters) to inform device-mismatch models; and aging/thermal effects are typically secondary at our scales and can be modeled as additional parametric variations.

\subsection{Yield Analysis Platform}
\label{ssec:baseline_algorithms}

OpenYield provides optimized implementations of established yield analysis methods, creating standardized benchmarks for algorithm comparison and reproducible research.

\subsubsection{Monte Carlo Reference}
Standard Monte Carlo is widely regarded as the ``gold standard" for yield estimation due to its unbiased nature and guaranteed convergence as the number of samples increases, albeit at the cost of high computational effort for rare-event problems~\cite{shi2018fast,xing2024every,dolecek2008breaking}. 
This method serves as the accuracy baseline, automating variation sampling, SPICE simulation, and pass/fail evaluation. Enhanced with stratified sampling and control variates, the implementation reduces required samples by $3$-$5\times$ while maintaining unbiased estimation—essential for validating advanced methods.

\subsubsection{Importance Sampling Variants}
Importance Sampling (IS) and its variants are considered robust alternatives for yield analysis: they require fewer parameters and hyperparameters than surrogate-based approaches, and they reliably converge for a wide range of yield estimation problems~\cite{dolecek2008breaking,shi2018fast,xing2024every}. 
Four IS methods address different yield analysis challenges. Mean-shifted IS (MNIS) \cite{dolecek2008breaking} translates distributions toward the most probable failure point: $\boldsymbol{\mu}_{shift} = \mathop{\arg\min}\limits_{\mathbf{x}} \|\mathbf{x}\|^2$ subject to failure boundary $f(\mathbf{x}) = 0$. This achieves $100\times$ speedup when single modes dominate, particularly for read failures driven by sense amplifier mismatch.
Adaptive Compressed Sampling (ACS) \cite{shi2019adaptive} builds sparse failure representations using compressed sensing, reducing samples by 90\% for smooth failure boundaries. Adaptive IS (AIS) \cite{shi2018fast} iteratively refines proposal distributions through cross-entropy minimization, automatically discovering multiple failure modes. High-dimensional Sparse Compressed Sampling (HSCS) \cite{wu2016hyperspherical} exploits sparsity to scale efficiently to hundreds of parameters, crucial for full-array analysis with systematic variations.

Surrogate model-based methods \cite{yin2022efficient} are not considered here, as they typically demand significant model training and hyperparameter tuning, making them more suitable for extremely complex, high-dimensional problems where MC is infeasible and IS-based methods may also fail.

\subsection{Standardized SRAM Optimization Platform}
\label{ssec:optimization}
OpenYield's optimization platform provides a standardized benchmarking environment for SRAM transistor sizing algorithms, enabling fair comparison of different optimization approaches under identical representative circuit conditions. The platform supports flexible optimization formulations including single-objective, multi-objective, and constrained problems, allowing users to combine different objectives and constraints based on their specific requirements.

\subsubsection{Flexible Optimization Formulation}

The platform accommodates diverse optimization scenarios through configurable objective functions and constraints. For demonstration, we present a single-objective formulation using a 6T SRAM cell within a $32 \times 2$ array configuration. The optimization explores design parameters $\mathbf{x}\in \mathbb{R}^{6}$: device types with different threshold voltages (\texttt{vtl}/\texttt{vtg}/\texttt{vth}), gate widths (\texttt{W}) of M0/M2/M4, and gate lengths (\texttt{L}). Parameters \texttt{W} and \texttt{L} range from 0.5 to 1.5$\times$ nominal values with 5nm steps. The optimization problem can be formulated as:
\begin{equation}
\label{eq:problem}
\begin{split}
\mathop{\arg\min}\limits_{\mathbf{x}} \left (\text{FoM}\right ), \text{s.t.}~T_{\text{READ}} \leq 0.5\text{ns}, T_{\text{WRITE}} \leq 0.5\text{ns},
\end{split}
\end{equation}
where 
\begin{equation}
\text{FoM} = \log_{10}\left(\frac{\mathrm{min(HSNM, RSNM, WSNM)}}{\mathrm{max}(P_{\text{READ}}, P_{\text{WRITE}}) \times \sqrt{\mathrm{area}}}\right).
\end{equation}
Users can easily customize objectives (power, area, noise margins), constraints (timing, yield requirements), and combine multiple formulations to address specific design challenges under representative circuit conditions.

\subsubsection{Integrated Optimization Algorithms}

The platform implements five state-of-the-art optimization algorithms for comprehensive benchmarking: Constrained Bayesian Optimization (CBO)~\cite{cbo2014} using Gaussian Process (GP) surrogates with Expected Hypervolume Improvement (EHI) acquisition; Particle Swarm Optimization (PSO)~\cite{pso1995} with adaptive parameters for population-based search; Simulated Annealing (SA)~\cite{sa1983} with exponential cooling schedules; Sequential Model-based Algorithm Configuration (SMAC)~\cite{smac2022} for automated hyperparameter optimization; and RoSE-Opt~\cite{cao2024rose} combining Bayesian optimization with reinforcement learning for adaptive exploration.

Each algorithm operates under identical evaluation budgets and circuit conditions, enabling fair performance comparison across different optimization paradigms.
The platform provides a unified evaluation framework where all algorithms utilize OpenYield's higher-fidelity circuit models. Each optimization run performs an identical number of SPICE simulations, ensuring fair computational comparison. Comprehensive metrics track convergence characteristics, Pareto front quality, and final design performance. The platform's integration with OpenYield's higher-fidelity models ensures optimized designs account for the complex failure mechanisms dominating manufactured arrays, while enabling reproducible comparison of optimization algorithm effectiveness.

\section{Experimental Results}
\label{sec:exp_results}

We organize this section around four questions: scalability and expected access-delay/power scaling (\autoref{ssec:exp_generator_scalability}); impact of second-order effects (\autoref{ssec:exp_second_order_impact}); agreement of baseline yield estimators with MC on realistic circuits at lower cost (\autoref{ssec:exp_algo_validation}); benefits of optimizing on SRAM cells under constraints (\autoref{ssec:exp_design_optimization}). 

All SRAM circuits for this evaluation were generated using the OpenYield framework. To ensure reproducibility, we primarily targeted a generic FreePDK45nm Technology Model \cite{FreePDK45}. The nominal supply $\texttt{VDD}=1.0\text{V}$. The default design parameters (all \texttt{W}s and \texttt{L}s refer to those in OpenRAM \cite{guthaus2016openram}).
Process parameters and their variations, including threshold voltage without bias (\texttt{vth0}), low-field mobility (\texttt{\textmu 0}), and offset voltage in the subthreshold region (\texttt{voff}), were chosen based on typical values with $\sigma=5\%$. Computationally intensive SPICE simulations were executed through a parallel version of Xyce\cite{keiter2013xyce} to ensure timely completion.

\begin{figure}[!tb] 
\centering
\begin{subfigure}[b]{0.49\linewidth} 
\setlength{\abovecaptionskip}{0.4pt}
\includegraphics[width=\textwidth]{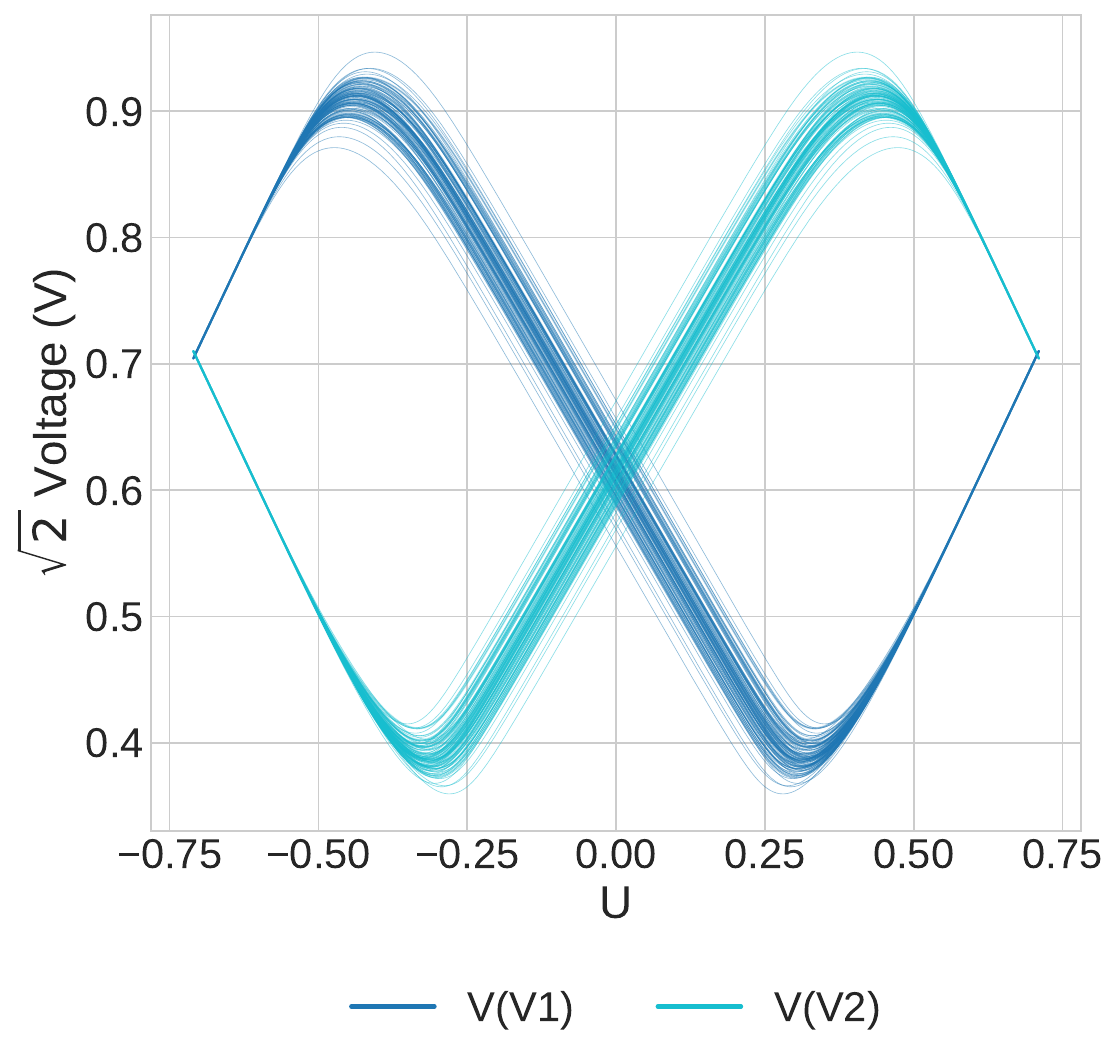} 
\caption{Hold SNM.}
\label{fig:hsnm_waveform}
\end{subfigure}
\hfill 
\begin{subfigure}[b]{0.49\linewidth}
\setlength{\abovecaptionskip}{0.4pt}
\includegraphics[width=\textwidth]{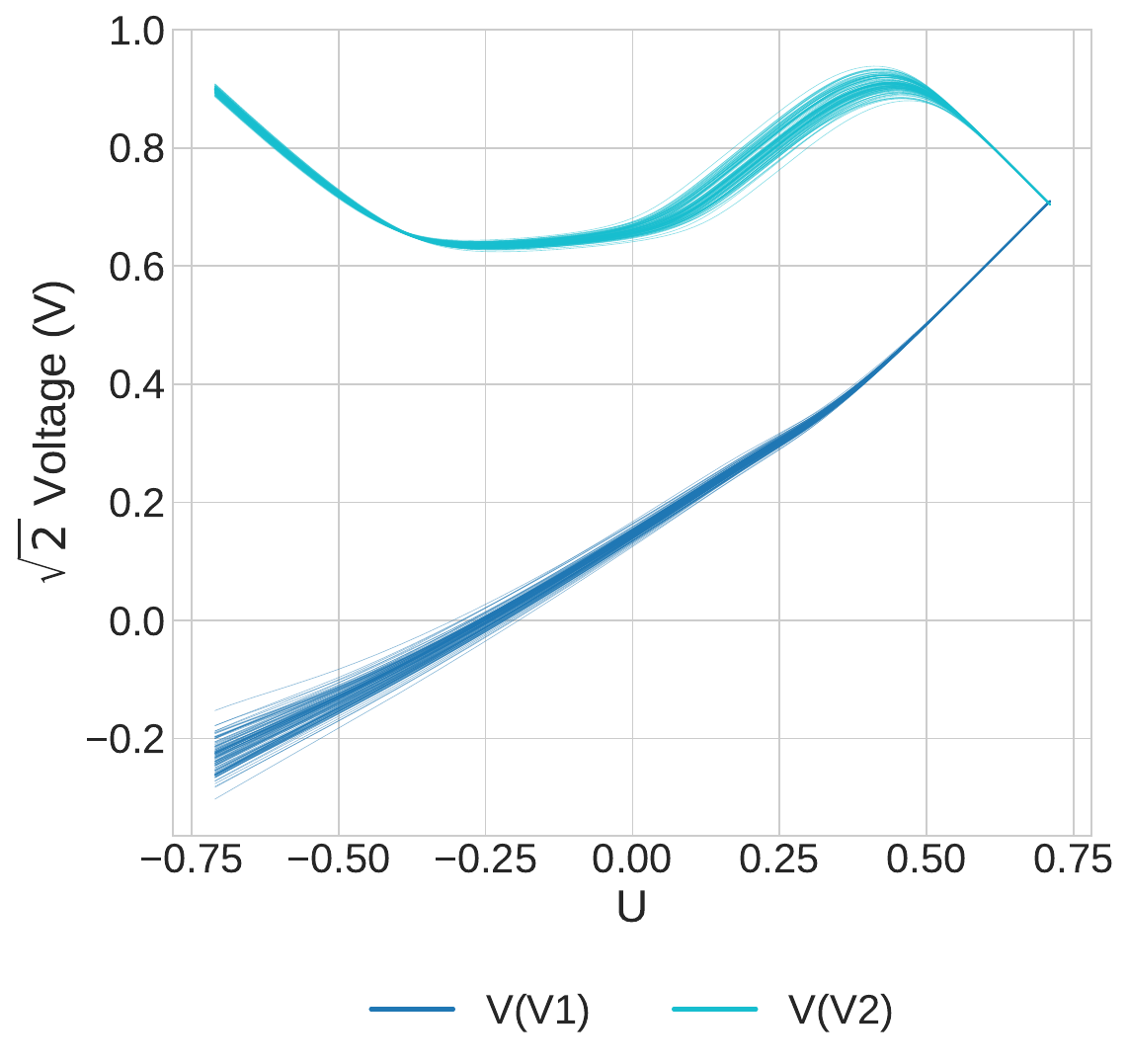} 
\caption{Write SNM.}
\label{fig:wsnm_waveform}
\end{subfigure}
\caption{Butterfly curves using DC analysis. The waveforms have been rotated by 45$^\circ$ as SNM is defined by the largest vertical distance between \texttt{V1} and \texttt{V2}.}
\label{fig:snms} 
\vspace{-4pt}
\end{figure}

Our primary experimental metrics are explained as follows. \textit{Yield}, $1 - P_f$, where $P_f$ is the failure probability from unmet operational criteria (e.g., timing metrics). 
\textit{Static Noise Margins} (SNM), including RSNM, WSNM, and HSNM, characterized via standard methods (e.g., largest square embedded within the lobes of butterfly curves) \cite{Kanj2009statistical, agarwal2008statistical}. 
\autoref{fig:hsnm_waveform} presents the HSNM curve. The outputs of the left- and right-side of inverters are denoted as V(\texttt{V1}) and V(\texttt{V2}) in \autoref{fig:snms}. We rotate all SNM curves by 45$^\circ$ to explicitly depict the largest noise margin. A typical RSNM butterfly curve has a similar shape to HSNM (waveform is omitted). Note that the RSNM value is always smaller than HSNM due to \texttt{BL} or \texttt{BLB} cannot be ideally high or low due to the turned-on access transistors.
\autoref{fig:wsnm_waveform} illustrates the write margin. The asymmetry of WSNM is caused by the write drivers. More space between the 2 curves makes it easier to write this cell.

\begin{figure}[!tb] 
\centering
\begin{subfigure}[b]{0.48\linewidth}
\setlength{\abovecaptionskip}{0pt}
\includegraphics[width=\textwidth]{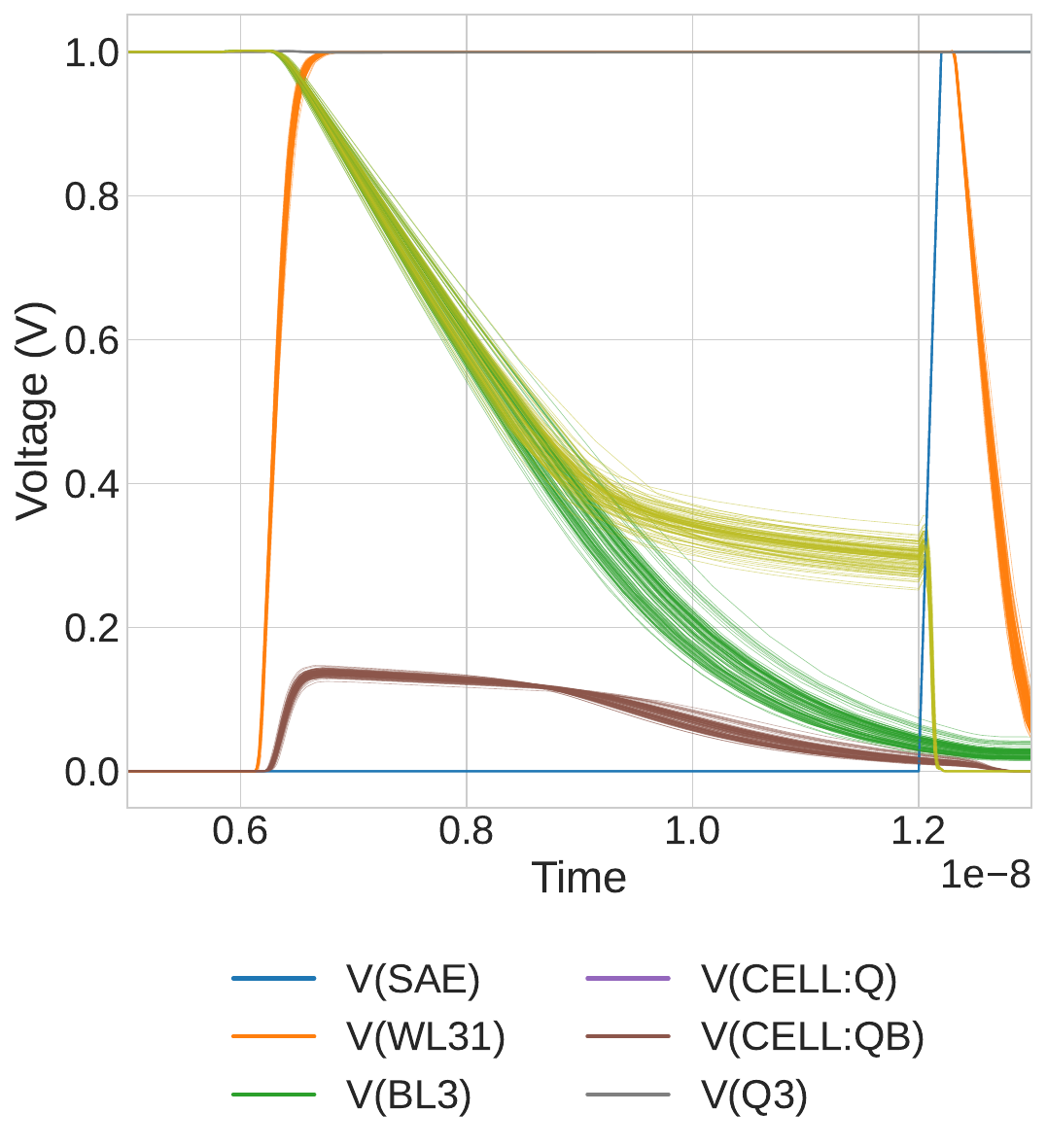} 
\caption{Read operation.}
\label{fig:read_waveform}
\end{subfigure}
\hfill 
\begin{subfigure}[b]{0.49\linewidth}
\setlength{\abovecaptionskip}{0.5pt}
\includegraphics[width=\textwidth]{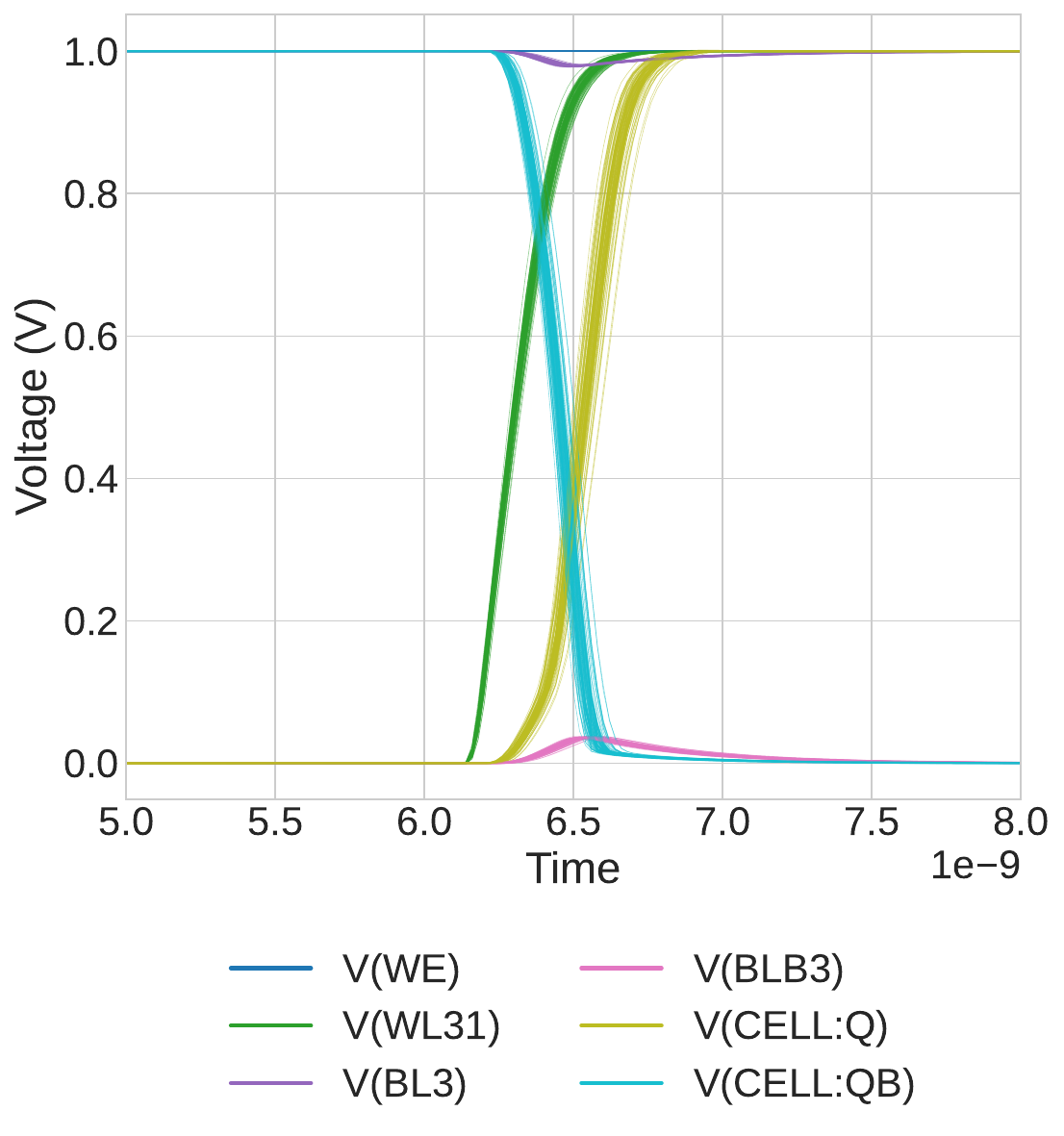} 
\caption{Write operation.}
\label{fig:write_waveform}
\end{subfigure}
\caption{Read and write operation under process variation.}
\label{fig:access} 
\vspace{-4pt}
\end{figure}

\textit{Read Time} ($T_{\text{READ}}$), from wordline (\texttt{WL}) assertion to a sufficient bitline (\texttt{BL}/\texttt{BLB}) differential for sense amplifier reliability. \autoref{fig:read_waveform} depicts a typical read cycle of a 32$\times$4 SRAM array with full peripheral circuits. Initially, V(\texttt{BL3}) and V(\texttt{BLB3}) are precharged high. Upon assertion of V(\texttt{WL31}), the cell begins to discharge \texttt{BL3} due to V(\texttt{CELL:Q3})$=0$. The time taken to develop a differential voltage is a key component in $T_{\text{READ}}$. After a sufficient voltage differential is established (e.g., 250mV), \texttt{SAE} is assertion to activate the sense amplifier, converting the analog voltage swing to a logical level.
\textit{Write Time} ($T_{\text{WRITE}}$), from \texttt{WL} assertion and active \texttt{BL}/\texttt{BLB} drive until internal storage nodes successfully flip (V(\texttt{CELL:Q})$=90\%\texttt{VDD}$). A representative write cycle is shown in \autoref{fig:write_waveform}. After \texttt{BL3} and \texttt{BLB3} are driven to complementary target values, the write operation is initiated by asserting Write Enable (\texttt{WE}) and pulling up \texttt{WL31}. A transition state of \texttt{CELL:Q/QB} kicks in, and finally, they reach a new stable state. Typically, $T_{\text{WRITE}}$ is much shorter than $T_{\text{READ}}$ in a balanced SRAM design, this is because (1) shorter interconnection of \texttt{CELL:Q/QB} has smaller parasitics than that of the read port (i.e., \texttt{BL}s); (2) the PU MOS (M4/M5) is designed to be weaker to increase the write stability. 

Other performance metrics include power consumption, covering static and dynamic power, and cell area (calculated based on the DRC rules in FreePDK45).

\subsection{Experiment 1: Scalability of the OpenYield Generator}
\label{ssec:exp_generator_scalability}

This subsection demonstrates the versatility of the OpenYield generator by showcasing its ability to produce SRAM arrays of varying sizes and complexities. 
%
\begin{figure}[!tb]
\centering
\begin{subfigure}[b]{0.49\linewidth}
\setlength{\abovecaptionskip}{0.8pt}
\includegraphics[width=\textwidth]{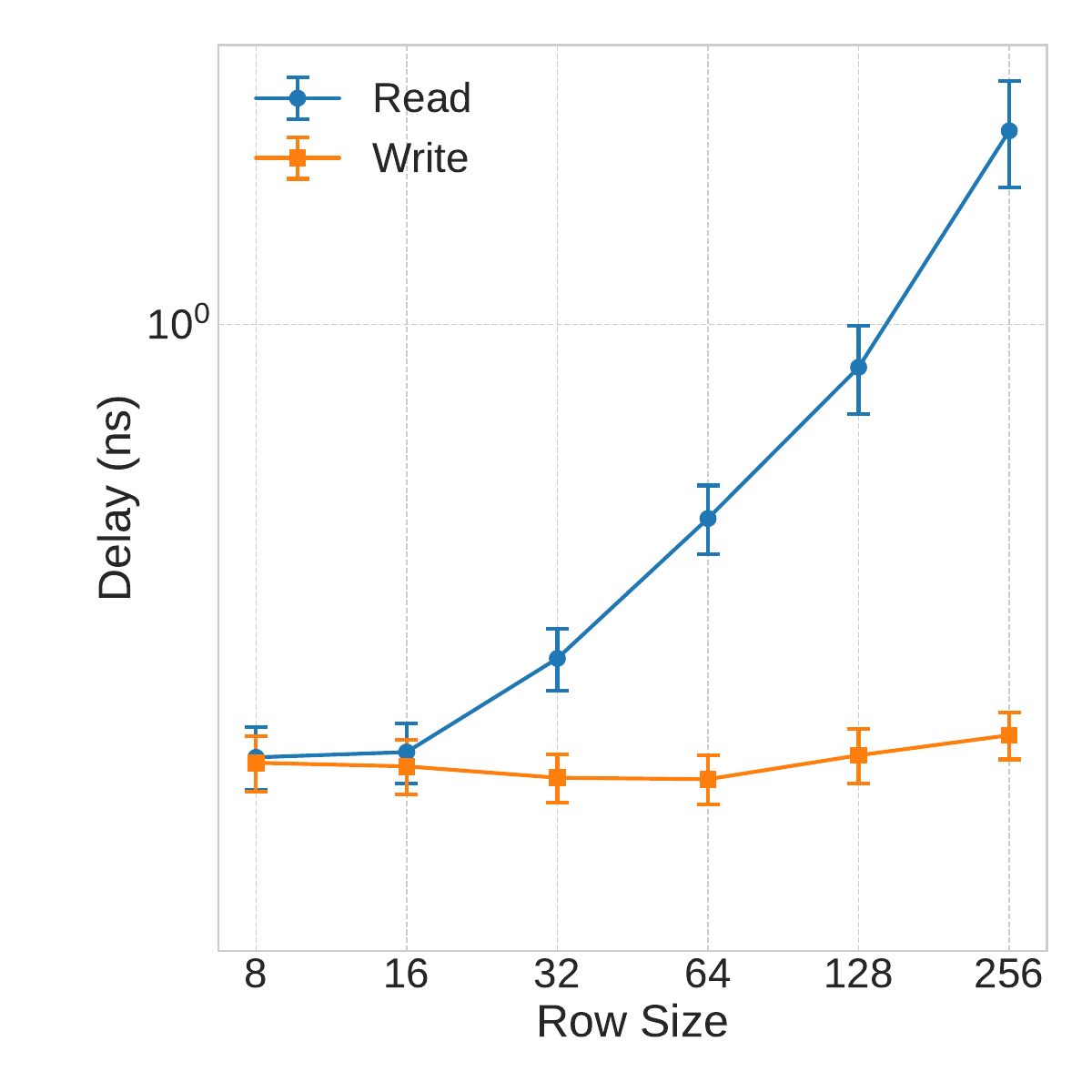} 
\caption{Mean access delay.} 
\label{fig:delay-vs-row}
\end{subfigure}
\hfill 
\begin{subfigure}[b]{0.49\linewidth}
\setlength{\abovecaptionskip}{0.8pt}
\includegraphics[width=\textwidth]{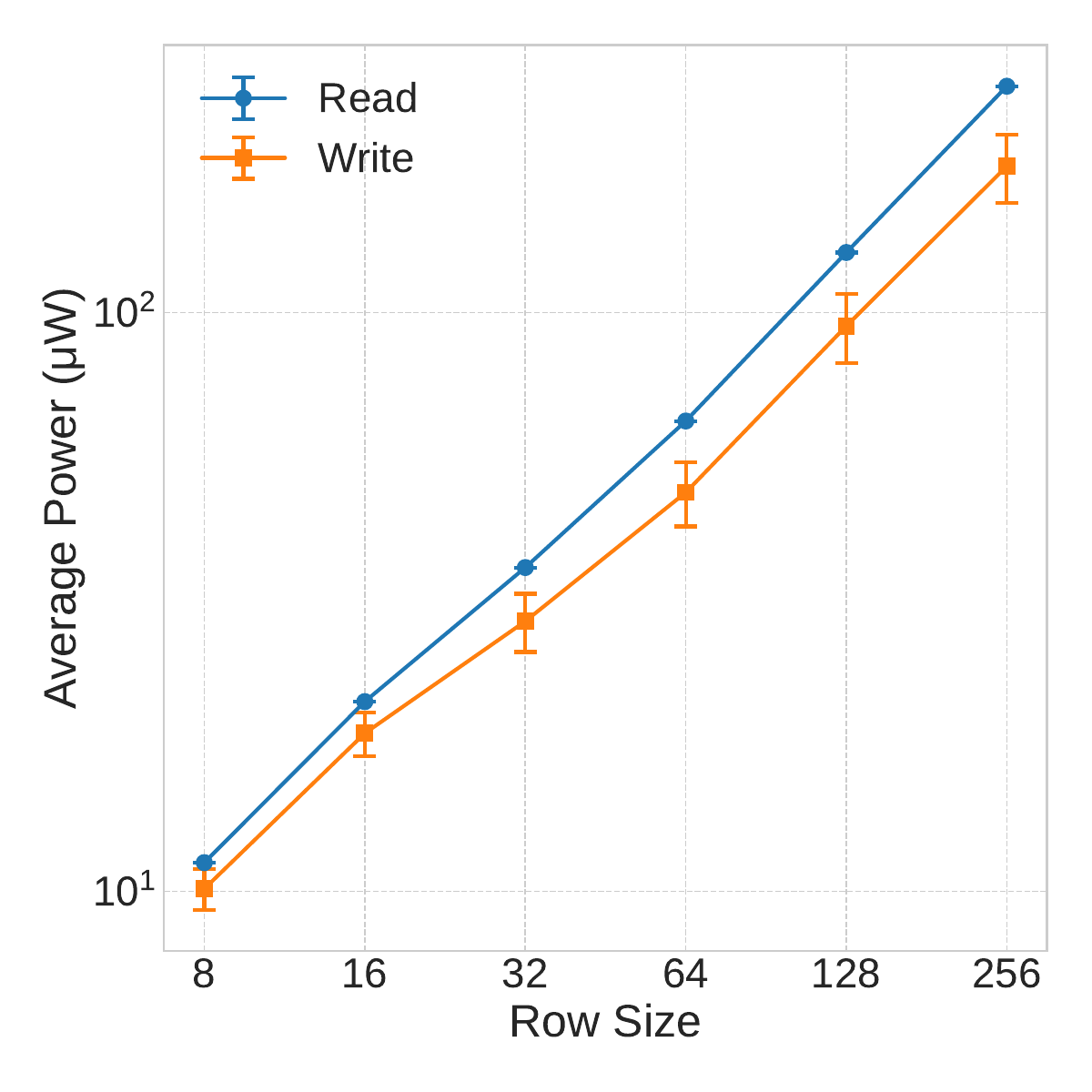} 
\caption{Average power.} 
\label{fig:power-vs-row}
\end{subfigure}
\caption{Mean read/write access delay and average read/write power consumption for a 4-column SRAM array as a function of the number of rows (8 to 256).} 
\label{fig:perf-vs-row} 
\vspace{-4pt}
\end{figure}
\autoref{fig:perf-vs-row} illustrates the impact of scaling the number of rows on key performance metrics for a 4-column SRAM array. Specifically, it shows the mean access delay and average access power as the number of rows varies from 8 to 256. The data indicates a clear trend: $T_{\text{READ}}$ increases from approximately 0.33ns for an 8-row array to 1.64ns for a 256-row array. Similarly, the average $P_{\text{READ}}$ during read operations scales from 11.2\textmu W to 246.1\textmu W across the same range of row sizes. 
In contrast, $T_{\text{WRITE}}$ shows minimal variation with the number of rows. $P_{\text{WRITE}}$, while also scaling with array size (10.1\textmu W $\rightarrow$ 179.0\textmu W), is substantially lower than read power.

Read delay and power grow strongly with array depth, while write metrics vary little; OpenYield's generator produces consistent arrays that expose these expected scaling trends.

\subsection{Experiment 2: Quantifying Second-Order Effects}
\label{ssec:exp_second_order_impact}

To quantify the impact of a key second-order effect, we performed comparative MC simulations on SRAM arrays of varying row sizes. One set of simulations included detailed parasitic resistance and capacitance (RC) models for interconnects, while the other intentionally omitted them, representing an idealized scenario. We focused these comparisons on read operations, as read delay and stability are often the most critical performance limiters in SRAM design, as mentioned before.

\subsubsection{Impact of Interconnect Parasitics} 
The inclusion of parasitic RC elements associated with the interconnects is expected to substantially affect both access delay and power consumption, particularly as array dimensions increase.
\begin{figure}[!tb]
\centering
\begin{subfigure}[b]{0.49\linewidth}
\setlength{\abovecaptionskip}{0.5pt}
\includegraphics[width=\textwidth]{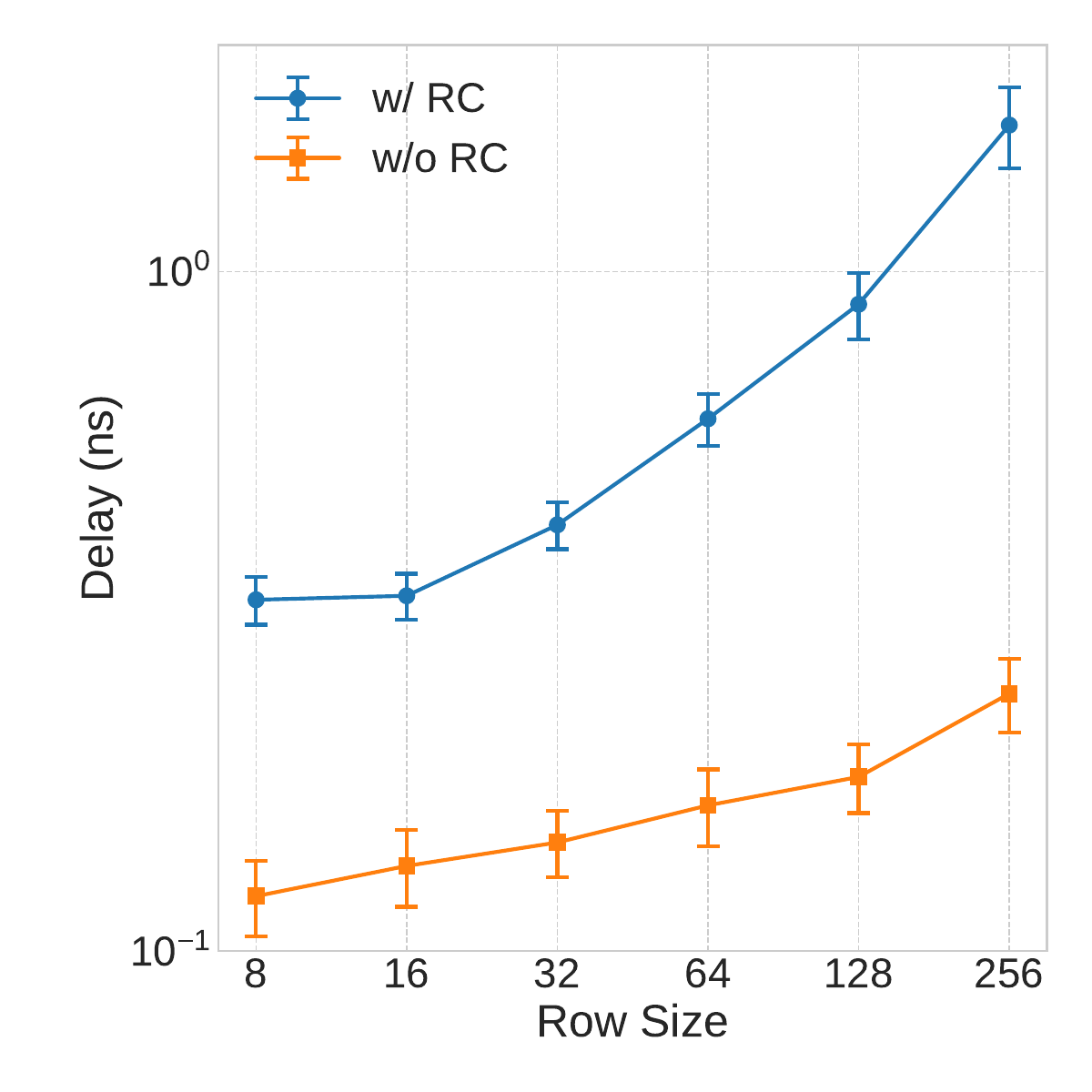} 
\caption{Mean and std of $T_{\text{READ}}$.} 
\label{fig:rc-delay-vs-row}
\end{subfigure}
\hfill 
\begin{subfigure}[b]{0.49\linewidth}
\setlength{\abovecaptionskip}{0.5pt}
\includegraphics[width=\textwidth]{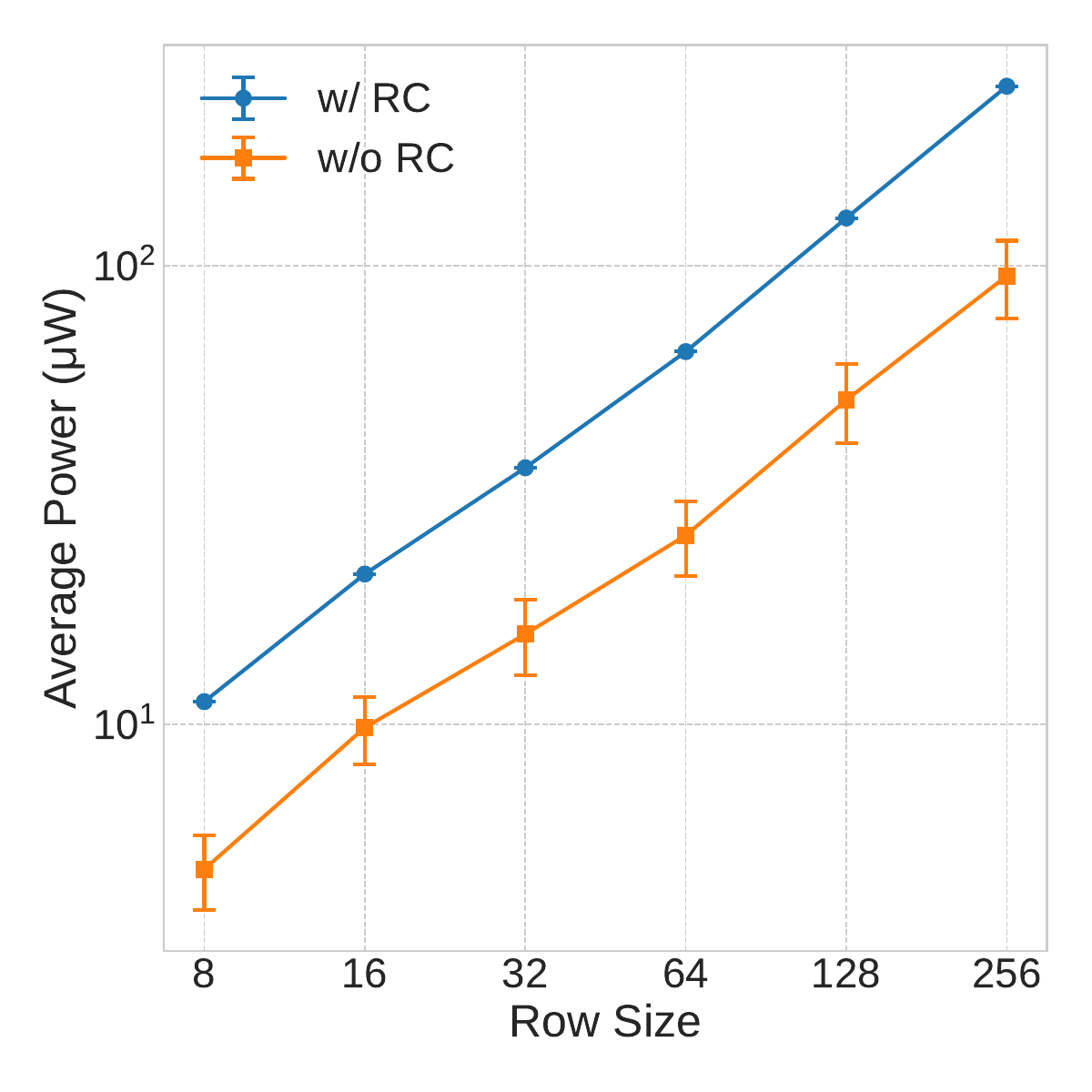} 
\caption{Mean and std of $P_{\text{READ}}$.} 
\label{fig:rc-power-vs-row}
\end{subfigure}
\caption{Impact of interconnect RC parasitics on mean read access delay and average read power consumption for 4-column SRAM arrays with varying numbers of rows.} 
\label{fig:parasitics-impact} 
\vspace{-4pt}
\end{figure}
The results presented in \autoref{fig:parasitics-impact} clearly demonstrate the profound impact of interconnect parasitics. For instance, for a 64-row array, the mean read access delay increases from approximately 0.164ns without parasitics to 0.607ns when parasitics are included—an increase by a factor of roughly 3.7. This discrepancy becomes even more pronounced for larger arrays; for a 256-row array, the delay with parasitics is approximately 6.87$\times$ greater than without.
Similarly, the average $P_{\text{READ}}$ is significantly underestimated when parasitics are ignored. For the 64-row array, including parasitics increases power by $~2.51 \times$. For the 256-row array, $P_{\text{READ}}$ rises by $2.59\times$. 

\subsubsection{Impact of Inter-cell Leakage (Leakage Noise)}
To characterize the impact of this leakage noise, we simulated a 64$\times$4 SRAM array across a range of supply voltages (\texttt{VDD} from 0.45V to 1.0V). We investigated the read delay under two scenarios representing different leakage conditions from idle cells in the same column: (1) all idle cells storing `0', and (2) all idle cells storing `1'. 

\begin{figure}[!tb] 
\centering
\includegraphics[width=0.9\linewidth]{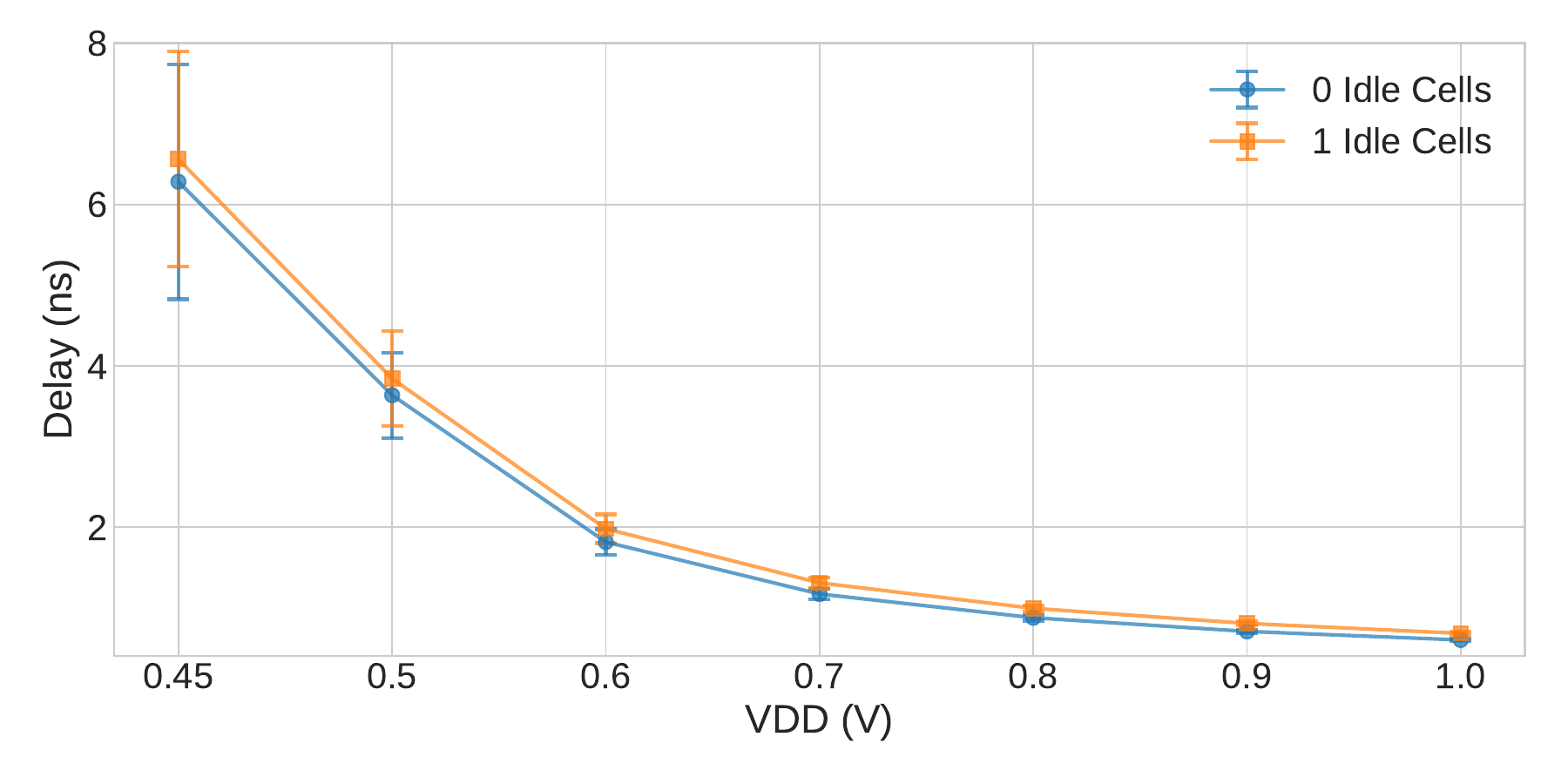} 
\caption{Read delay comparison for a 64$\times$4 SRAM array under different inter-cell leakage conditions (idle cells storing `0' vs. `1') across various supply voltages (\texttt{VDD}).
}
\label{fig:leak-vs-vdd}
\vspace{-4pt}
\end{figure}

As illustrated in \autoref{fig:leak-vs-vdd}, 
Our first finding is that the read access delay is strongly dependent on the supply voltage, decreasing from over 6ns at $\texttt{VDD}=0.45$V to approximately 0.84ns at $\texttt{VDD}=1.0$V for both idle-cell data patterns. 
At lower supply voltages, the effect of leakage noise is more pronounced. For example, at $\texttt{VDD}=0.45$V, the read delay when idle cells store `0' (6.28ns) is about 0.28ns (or $\approx 4.5\%$) shorter than when idle cells store `1' (6.56ns). However, this difference diminishes as \texttt{VDD} increases, becoming negligible at $\texttt{VDD}=0.7$V and above. 

\subsubsection{Impact of Peripheral Circuit Variations}
The variations of peripheral circuits play a crucial role in overall SRAM performance and reliability. To demonstrate their impact, we compare simulations performed with a simplified model versus the model with full peripherals for a 64$\times$32 SRAM array. 

\autoref{tab:delay_power_comparison_grouped} summarizes key delay and power metrics for read and write operations under these two modeling conditions.
\begin{table}[tbp]
\centering
\caption{Comparison of Delay and Power Metrics with and without Peripheral Circuits. Array Size = 64$\times$32.}
\label{tab:delay_power_comparison_grouped}
    \begin{tabular}{lcc|cr} 
        \toprule
        \textbf{R/W} & \textbf{Type} & \textbf{Metric} & \textbf{w/o Peri.} & \textbf{w/ Peri.} \\
        \midrule
        \multirow{7}{*}{\rotatebox[origin=c]{90}{\textbf{Read Operation}}} 
        & \multirow{4}{*}{Delay (ns)}
            & $T_\text{DEC}$ & N/A & 0.0962\\
        &   & $T_\text{PRECH}$ & N/A & 0.1597 \\
        &   & $T_\text{WLDRV}$ & N/A & 0.0759 \\
        &   & $T_\text{BL}$    & 0.2906 & 0.3378 \\
        \cmidrule{2-5} 
        & \multirow{3}{*}{Power (mW)}
            & avg$(P_{\text{READ}})$   & 0.4404 & 0.5649 \\
        &   & dyn$(P_{\text{READ}})$  & 0.6291 & 0.8354 \\
        &   & static$(P_{\text{READ}})$ & 0.0061 & 0.3378 \\
        \midrule 
        \multirow{7}{*}{\rotatebox[origin=c]{90}{\textbf{Write Operation}}} 
        & \multirow{4}{*}{Delay (ns)} 
            & $T_\text{DEC}$ & N/A & 0.0922 \\
        &   & $T_\text{WDRV}$  & N/A & 0.0971 \\
        &   & $T_\text{WLDRV}$ & N/A & 0.0746 \\
        &   & $T_\text{Q}$ & 0.0378 & 0.0556 \\
        \cmidrule{2-5} 
        & \multirow{3}{*}{Power (mW)}
            & avg$(P_{\text{WRITE}})$   & 0.0129 & 0.2453 \\
        &   & dyn$(P_{\text{WRITE}})$ & 0.0371 &  1.5994\\
        &   & static$(P_{\text{WRITE}})$   & 0.0047 & 0.0168\\
        \bottomrule
    \end{tabular}
\vspace{-2pt}
\end{table}
%

For read operations, critical delay components associated with peripherals like precharge time and wordline drive time are inherently captured. Furthermore, the read delay ($T_\text{DEC}+T_\text{PRECH}+T_\text{WLDRV}+T_\text{BL}$) shows an increase of {$2.3\times$ with peripherals (0.2906ns $\rightarrow$ 0.6696ns)}. This highlights the considerable loading and delay contributions from components like sense amplifiers and column multiplexers. Read power consumption ($P_{\text{READ}}$) is also markedly higher with peripherals: average $P_{\text{READ}}$ increases by {$1.28\times$, and dynamic $P_{\text{READ}}$ by $1.33\times$}.

For write operations, similar trends are observed. Delays specific to write drivers and wordline drivers are accounted for with the full model. The write time ($T_\text{DEC}+T_\text{WDRV}+T_\text{WLDRV}+T_\text{Q}$) also has a $\approx 8.45\times$ increase when peripheral interactions are considered {(0.0378ns $\rightarrow$ 0.3194ns)}. Write power ($P_{\text{WRITE}}$) metrics also show increases: average $P_{\text{WRITE}}$ is {$\approx 19.02\times$ higher, dynamic $P_{\text{WRITE}}$ is $\approx 43.23\times$ higher, and notably, static $P_{\text{WRITE}}$ sees an increase of over $3.57\times$} when peripheral circuits are included.

These comparisons underscore that omitting detailed models of peripheral circuits and their variations can lead to highly optimistic and inaccurate estimations of SRAM performance, particularly for delay and power. 

\subsection{Experiment 3: Validation of Baseline Yield Algorithms}
\label{ssec:exp_algo_validation}

This experiment validates the baseline yield analysis algorithms in OpenYield, focusing on Standard Monte Carlo (MC) and Importance Sampling (IS) methods. 
The test circuits include: (i) a single 6T cell with 18 process variation parameters; (ii) a 3$\times$2 array with 108 parameters; and (iii) a 32$\times$2 array with 1152 parameters. Write delay serves as the performance metric across all experiments. The figure of merit is $\mathrm{std}(P_f) / P_f$, where $\mathrm{std}(P_f)$ represents the standard deviation of the estimated failure rate, serving as the termination criterion.

\begin{figure}
\vspace{-3pt}
    \centering
    \includegraphics[width=1.0\linewidth]{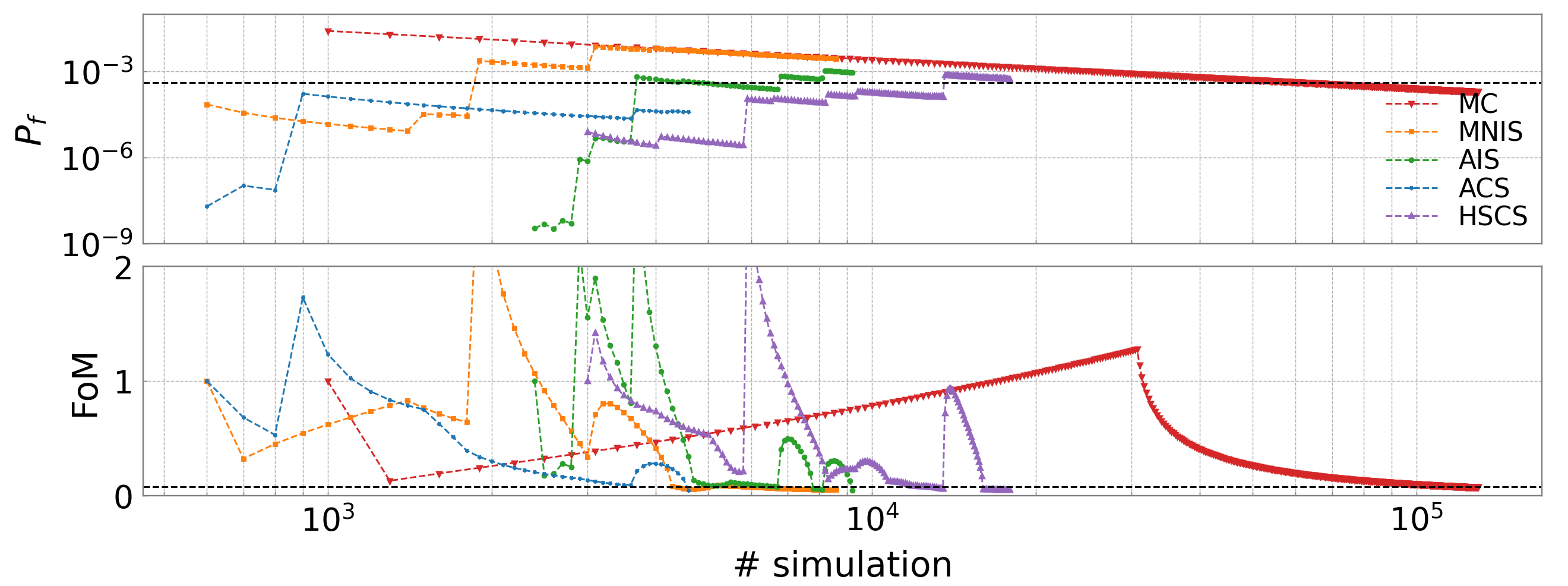}
    \caption{Failure estimation with FoM for 3$\times$2 SRAM array.}
    \label{fig:yield_estimation for 6-bit 6T-SRAM}
    \vspace{-0.2cm}
\end{figure}
\begin{figure}
    \centering
    \includegraphics[width=1.0\linewidth]{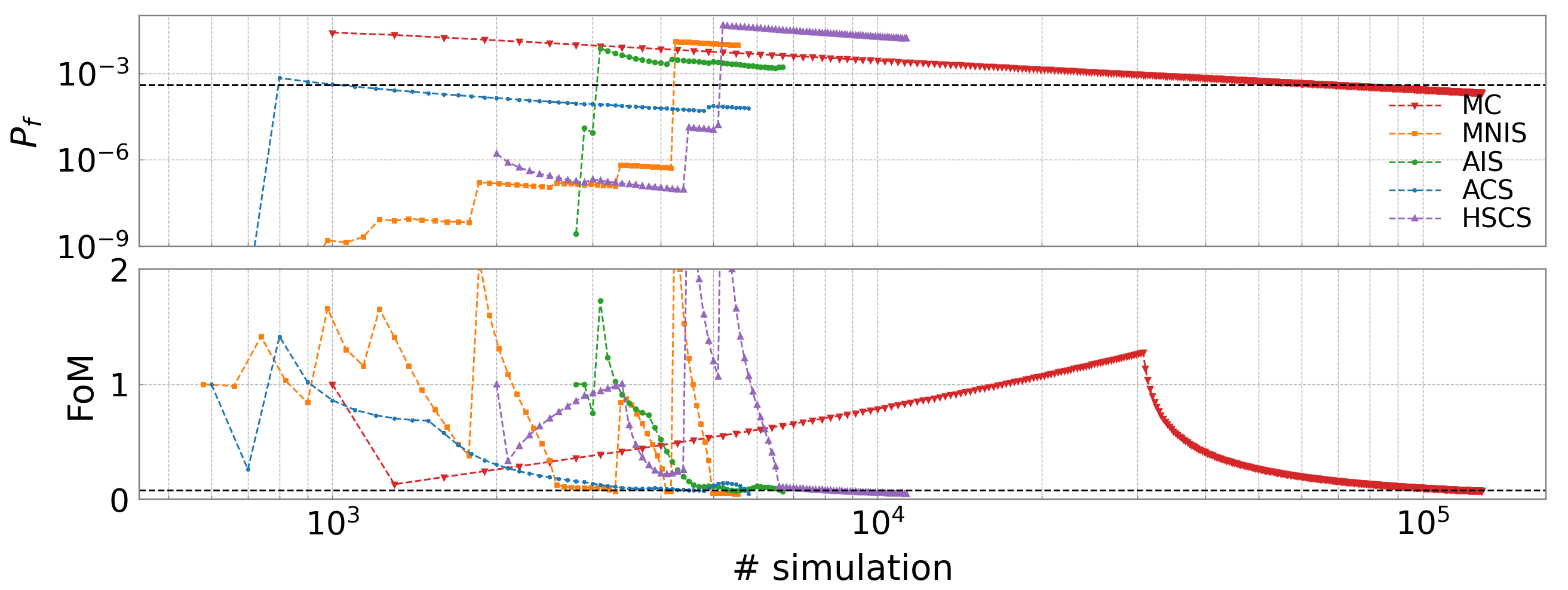}
    \caption{Failure estimation with FoM for 32$\times$2 SRAM array.}
    \label{fig:yield_estimation for 64-bit 6T-SRAM}
    \vspace{-4pt}
\end{figure}

\begin{table*}[t]
\centering
\caption{Comparison of Different Baseline Yield Analysis Methods on Various SRAM Circuits.}
\label{tab:yield_estimation result}
\begin{tabular}{l|ccrr|ccrr|ccrr}
\toprule
\multirow{2}{*}{\textbf{Method}} 
& \multicolumn{4}{c|}{\textbf{Cell}} 
& \multicolumn{4}{c|}{\textbf{\(3\times2\) Array}} 
& \multicolumn{4}{c}{\textbf{\(32\times2\) Array}} \\
\cmidrule(r){2-5} \cmidrule(r){6-9} \cmidrule(r){10-13}
& Fail & FoM & \#Sim & Time(s)
& Fail & FoM & \#Sim & Time(s)
& Fail & FoM & \#Sim & Time(s) \\
\midrule
\textbf{MC} & 2.4e-4 & 0.100 & 102,400 & 7,567
     & 2.0e-4 & 0.070  & 130,000 & 24,492 
     & 2.2e-4 & 0.076 & 128,500 & 199,447 \\
\textbf{MNIS}\cite{dolecek2008breaking} & 2.1e-4 & 0.017 & 12,450 & 1,075 
     & 2.7e-3 & 0.049  & 8,600 & 1,771 
     & 9.6e-3 & 0.049 & 5,540  & 10,178 \\
\textbf{AIS}\cite{shi2018fast} & 6.1e-4 & 0.056 & 10,050 & 934 
     & 9.3e-4 & 0.046 & 9,200 & 1,442 
     & 1.6e-3 & 0.069 & 6,700  & 7,091 \\
\textbf{ACS}\cite{shi2019adaptive}  & 1.7e-3 & 0.018 & 5,600 & 946 
     & 3.8e-5 & 0.049 & 4,600 & 1,035 
     & 6.2e-5 & 0.049 & 5,800  & 10,050 \\
\textbf{HSCS}\cite{wu2016hyperspherical} & 5.1e-2 & 0.049  & 11,900 & 1,469 
     & 5.5e-4 & 0.049 & 17,900 & 1,153 
     & 1.6e-2 & 0.049 & 11,300  & 10,312 \\
\bottomrule
\end{tabular}
\end{table*}

Results for the single 6T cell are shown in \autoref{tab:yield_estimation result}. MNIS~\cite{dolecek2008breaking} achieves near-MC accuracy with less than 10\% error while delivering 7$\times$ speedup. AIS~\cite{shi2018fast} provides a moderate balance between precision and efficiency, though with higher error than MNIS.
For the 3$\times$2 case (\autoref{fig:yield_estimation for 6-bit 6T-SRAM} and \autoref{tab:yield_estimation result}), ACS~\cite{shi2019adaptive} uniquely surpasses MC accuracy with sub-10\% error while achieving 23.7$\times$ speedup. However, MNIS exhibits significantly degraded performance with error rates 10$\times$ higher than MC, revealing limited capacity for complex problems. 
This behavior can be attributed to the MNIS approach, which shifts the sampling distribution toward the nearest failure regions to reduce the variance of the estimated failure rate. However, this strategy is only effective when there is a single failure region or when multiple failure regions are well separated. For more complex problems with multiple or overlapping failure regions, this approach may not yield optimal results \cite{xing2024every}.
AIS and HSCS, in contrast, show moderate precision as they are designed to deal with multiple failure regions and high-dimensional problems. However, they require substantially more simulations as a result of the need to explore the parameter space and to update their model parameters.
%
In the largest case, MC requires over 128,000 simulations and 55 hours of runtime. ACS achieves comparable accuracy with only 5,800 simulations—a 19.8$\times$ speedup while maintaining stable performance. HSCS achieves the fastest runtime but suffers a drastic accuracy loss, with failure estimates over an order of magnitude higher than MC. MNIS and AIS also display unstable results.

These results establish OpenYield as a transparent environment for benchmarking yield analysis algorithms, enabling reliable comparisons within the research community.


\subsection{Experiment 4: Design Improvement via Optimization} \label{ssec:exp_design_optimization}

\begin{figure*}
    \centering
    \begin{minipage}{0.74\linewidth}
        \centering
        \includegraphics[width=\linewidth]{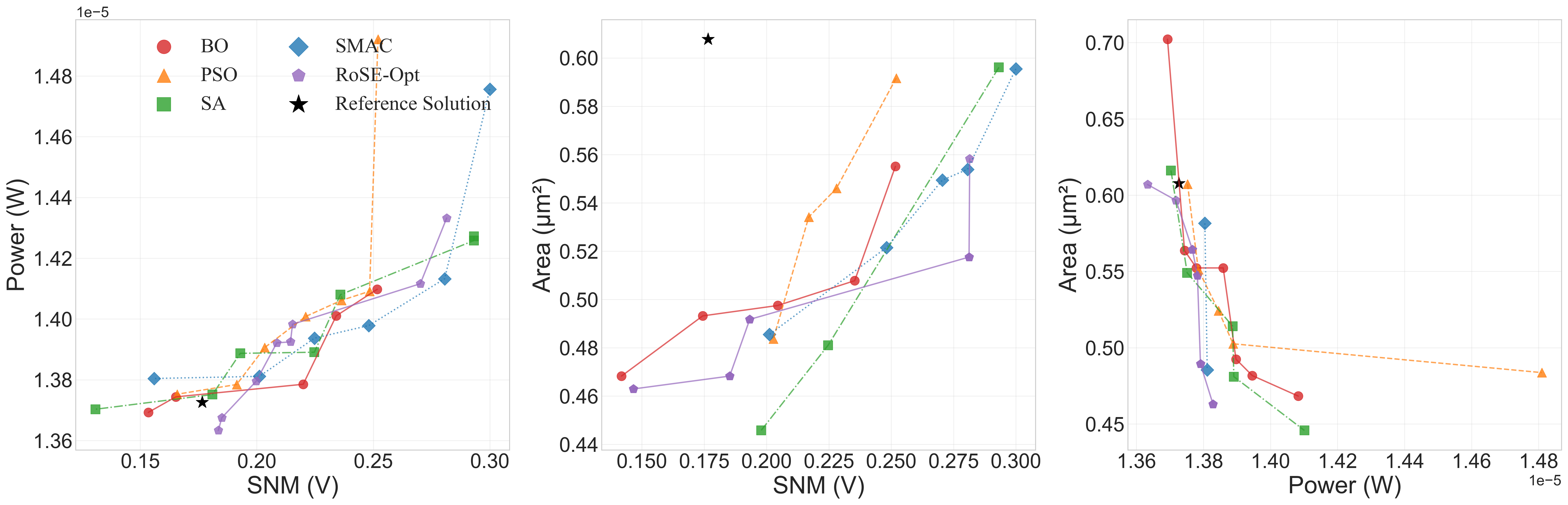}
        \caption{Pareto front of: (a) power vs. SNM; (b) area vs. SNM; (c) area vs. power.}
        \label{fig:pareto_front_results}
    \end{minipage}%
    \hfill
    \begin{minipage}{0.24\linewidth}
        \centering
        \includegraphics[width=\linewidth]{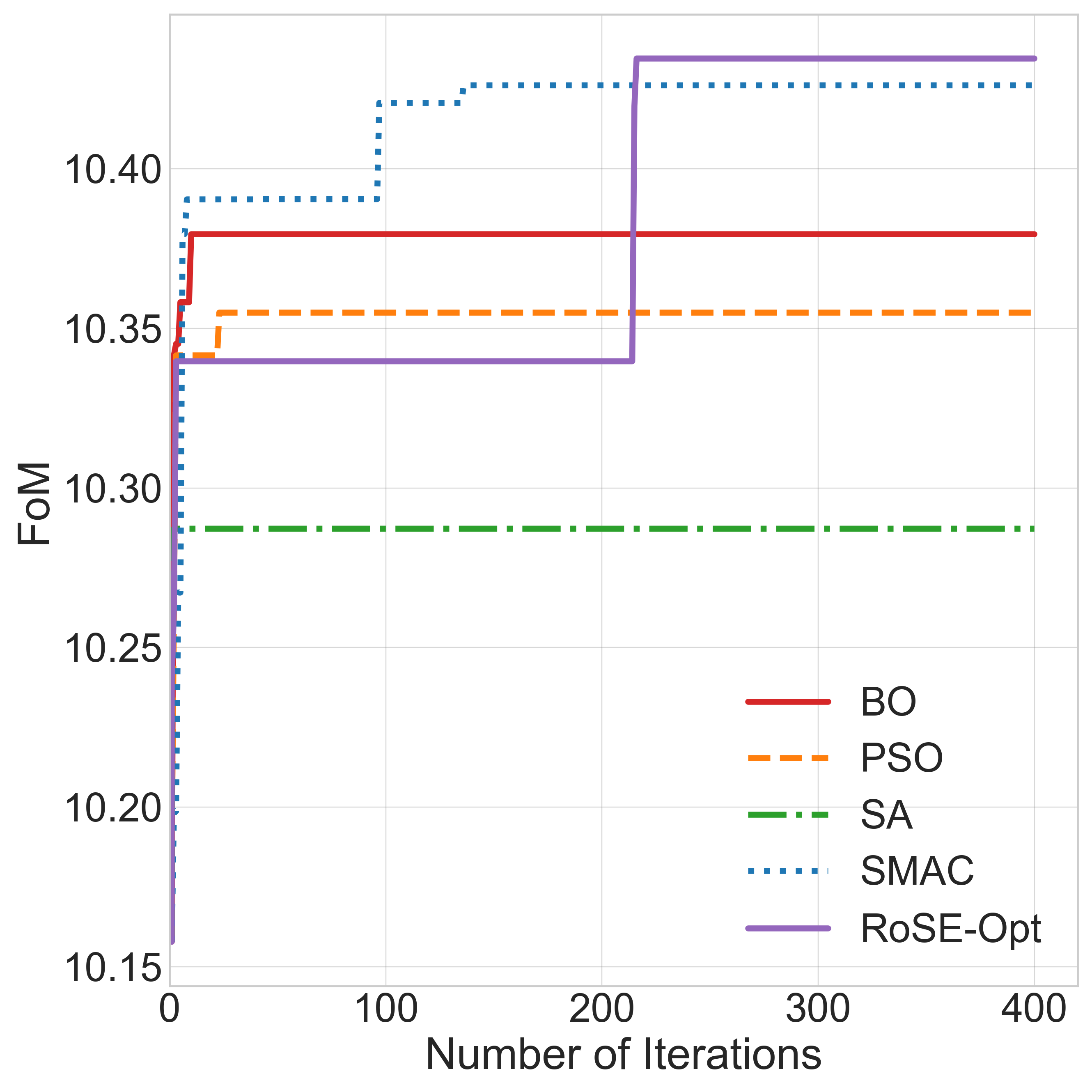}
        \caption{FoM comparison.}
        \label{fig:fom_results}
    \end{minipage}
    \vspace{-4pt}
\end{figure*}

In this sub-section, we focus on multi-objective transistor sizing optimization to improve cell robustness while maintaining performance constraints.
We implement a control study using simplified models (neglecting second-order effects) and re-evaluate the ``optimal'' design in \cite{guthaus2016openram}.
The optimization objective can be found in \autoref{eq:problem}.

We evaluate five state-of-the-art optimization algorithms, each performing 400 SPICE simulations: (i) \textit{CBO} \cite{cbo2014} using GP surrogates with EHI acquisition and constraint handling; (ii) \textit{PSO} \cite{pso1995} with population size 20, adaptive inertia weight ($w = 0.8$), and cognitive/social parameters ($c_1 = c_2 = 0.5$); (iii) \textit{SA} \cite{sa1983} with exponential cooling schedule ($T_0 = 1000$, $T_{\min} = 10^{-7}$, $\alpha = 0.98$); (iv) \textit{SMAC} employing sequential model-based configuration; and (v) \textit{RoSE-Opt} combining Bayesian optimization with reinforcement learning using Proximal Policy Optimization agents for adaptive parameter exploration.

\autoref{fig:pareto_front_results} shows the Pareto fronts and \autoref{fig:fom_results} compares convergence characteristics across all algorithms. \autoref{tab:optimization_impact} shows the OpenYield-optimized design (RoSE-Opt) achieves significant improvements: 65\% enhancement in read SNM (0.17V $\rightarrow$ 0.28V), 39\% improvement in write SNM (0.79V $\rightarrow$ 1.10V), and 15\% area reduction (0.61\textmu m$^2$ $\rightarrow$ 0.52\textmu m$^2$) while maintaining timing constraints. The composite FoM increases from 10.20 to 10.43, demonstrating balanced multi-objective improvement. SMAC and CBO show competitive performance with different exploration-exploitation trade-offs, while PSO and SA exhibit more stochastic convergence patterns.

Critically, designs optimized using simplified models exhibit poor performance when evaluated under higher-fidelity conditions, with some violating timing constraints or showing degraded stability margins. This validates the necessity of high-fidelity modeling during optimization, positioning OpenYield as an essential tool for robust SRAM design.

\begin{table}[tb]
\centering
\caption{Impact of Transistor Sizing Optimization on a 6T SRAM Cell.}
\label{tab:optimization_impact}
\begin{tabular}{ll|cc}
    \toprule
    \multicolumn{2}{c|}{\textbf{Parameter/Metric}} & \textbf{w/o Opt.} & \textbf{w/ RoSE-Opt} \\
    \midrule
    \multirow{6}{*}{\rotatebox[origin=c]{90}{\textbf{Design Params}}} & NMOS Model & NMOS\_VTG & NMOS\_VTH \\
    & Type & PMOS\_VTG & PMOS\_VTG \\
    & \texttt{W\_PD} (\textmu m) & 0.205 & 0.170 \\
    & \texttt{W\_PU} (\textmu m) & 0.090 & 0.124 \\
    & \texttt{W\_PG} (\textmu m) & 0.135 & 0.075 \\
    & \texttt{L} (nm) & 50 & 74 \\
    \midrule
    \multirow{9}{*}{\rotatebox[origin=c]{90}{\textbf{Performance Metrics}}} & $T_{\text{READ}}$ ($<$0.5ns) & 0.4533 & 0.4374 \\
    & $T_{\text{WRITE}}$ ($<$0.5ns) & 0.0903 & 0.0583 \\
    & $P_{\text{READ}}$ (\textmu W) & 13.74 & 14.37 \\
    & $P_{\text{WRITE}}$ (\textmu W) & 0.86 & 1.14 \\
    & HSNM (V) & 0.33 & 0.33 \\
    & RSNM (V) & 0.17 & 0.28 \\
    & WSNM (V) & 0.79 & 1.10 \\
    & Area (\textmu m$^2$) & 0.61 & 0.52 \\
    & FoM & 10.20 & 10.43 \\
    \bottomrule
\end{tabular}
\vspace{-6pt}
\end{table}


\section{Conclusion \& Future Work}
\label{sec:conclusion}
OpenYield is an open-source effort to narrow the gap between academic studies and practical SRAM analysis/optimization by providing accessible, higher-fidelity benchmarks and tooling that incorporate important second-order effects beyond common simplifications. We hope this resource enables more reproducible comparisons and techniques that transfer better toward real SRAM designs.

Future work includes: (i) support for multi-array and multi-bank organizations and their corresponding interfaces; (ii) incorporation of additional second-order effects; (iii) automatic layout generation; (iv) full design file generation to support physical design flow integration; and (v) integration of state-of-the-art yield estimation algorithms.
\bibliographystyle{IEEEtran}
\bibliography{refs}

\end{document}